\documentclass[twocolumn,journal]{IEEEtran}

\usepackage{amsfonts}
\usepackage{amsmath}
\usepackage{amsthm}
\usepackage{amssymb}
\usepackage{graphicx}
\usepackage{supertabular}
\usepackage{longtable}
\usepackage[usenames,dvipsnames]{color}
\usepackage{bbm}
\usepackage{fancyhdr}
\usepackage{breqn}
\usepackage{fixltx2e}
\usepackage{capt-of}
\usepackage{tikz}
\usetikzlibrary{matrix}
\usepackage{endnotes}
\usepackage{hyperref}
\usepackage{soul}
\usepackage{marginnote}

\newcommand{\mathsym}[1]{}
\newcommand{\unicode}[1]{}

\usepackage{soul}
\usepackage{float}

\usepackage{tabularx}

\usepackage[framemethod=TikZ]{mdframed}
\usepackage[framemethod=TikZ]{mdframed}
\usepackage{framed}

\mdfsetup{%
	outerlinewidth=1,skipabove=30pt,backgroundcolor=gray!10, outerlinecolor=black,innertopmargin=8pt,splittopskip=\topskip,skipbelow=\baselineskip, skipabove=\baselineskip,ntheorem,roundcorner=5pt}

\mdtheorem[nobreak=true,outerlinewidth=1,
backgroundcolor=yellow!50, outerlinecolor=black,innertopmargin=0pt,splittopskip=\topskip,skipbelow=\baselineskip, skipabove=\baselineskip,ntheorem,roundcorner=5pt,font=\itshape]{result}{Result}

\mdtheorem[nobreak=true,outerlinewidth=1,
backgroundcolor=yellow!50, outerlinecolor=black,innertopmargin=0pt,splittopskip=\topskip,skipbelow=\baselineskip, skipabove=\baselineskip,ntheorem,roundcorner=5pt,font=\itshape]{theorem}{Theorem}

\mdtheorem[nobreak=true,outerlinewidth=1,
backgroundcolor=gray!10, outerlinecolor=black,innertopmargin=0pt,splittopskip=\topskip,skipbelow=\baselineskip, skipabove=\baselineskip,ntheorem,roundcorner=5pt,font=\itshape]{remark}{Remark}
\mdtheorem[nobreak=true,outerlinewidth=1,
backgroundcolor=pink!30, outerlinecolor=black,innertopmargin=0pt,splittopskip=\topskip,skipbelow=\baselineskip, skipabove=\baselineskip,ntheorem,roundcorner=5pt,font=\itshape]{quaestio}{Quaestio}

\mdtheorem[nobreak=true,outerlinewidth=1,
backgroundcolor=yellow!50, outerlinecolor=black,innertopmargin=5pt,splittopskip=\topskip,skipbelow=\baselineskip, skipabove=\baselineskip,ntheorem,roundcorner=5pt,font=\itshape]{background}{Background}

\usepackage{hyperref}
\begin{document}
\title{{\color{Brown}
The Probability Conflation: A Reply }}

\author{{Nassim Nicholas Taleb\IEEEauthorrefmark{1},  Ronald Richman\IEEEauthorrefmark{2}, Marcos Carreira\IEEEauthorrefmark{3}, James Sharpe\IEEEauthorrefmark{4}\\
    \IEEEauthorblockA{ 
   \IEEEauthorrefmark{1}Tandon School of Engineering, New York University and Universa Investments\;
  \IEEEauthorrefmark{2}Old Mutual Insure and University of Witwatersrand\;
   \IEEEauthorrefmark{3}Ecole Polytechnique\;
   \IEEEauthorrefmark{4}Sharpe Actuarial Limited\\
  Forthcoming, \textit{International Journal of Forecasting}, 2023}
}
\thanks{\color{Brown}Corresponding author: N N Taleb, nnt1@nyu.edu. We thank Fergal McGovern and participants in the webinar Global Uncertainty Reading Group (formerly Global \textit{Technical Incerto} Reading Group) special meeting on Dec 2, 2022. We also thank Pasquale Cirillo, Sara Lapsley, Brandon Yarckin and Mark Spitznagel.} } 
 
 \maketitle 
 
 \begin{mdframed}
\begin{abstract}
	We respond to Tetlock et al. (2022) showing 1) how expert judgment fails to reflect tail risk, 2) the lack of compatibility between forecasting tournaments and tail risk assessment methods (such as extreme value theory).More importantly, we communicate a new result showing a greater gap between the properties of tail expectation and those of the corresponding probability.
\end{abstract}
\end{mdframed}

\section{The Fat Tails Problem}

 Tetlock et al. (2022) \cite{tetlock2022false}, in their criticism of claims by a paper titled "On single point forecasts for fat-tailed variables" in this journal (Taleb et al., 2022 \cite{taleb2020single}) insist that discriminating between a binary probability and a continuous distribution is a false dichotomy, that binary probabilities derived from expert forecasting tournaments can provide information on tail risk, in addition to some claims about a collaboration with the first author and a "challenge". 
 
 We apologize for not answering most of their points as these are already amply covered in two papers in this very journal, which includes the one they are criticizing, \cite{taleb2020single} and the more formal \cite{taleb2020differences}. Alas "probability" is a mathematical concept that does not easily accommodate verbal discussions and requires a formal treatment, which necessitates precise definitions.
 
 At the gist of what we referred to as "the so-called masquerade problem" is the following conflation, we simplify from \cite{taleb2020differences} using a continuous distribution for ease of exposition:

	  Let $K \in \mathbb{R}^+$ be a threshold, $f(.)$ a density function for  a random variable $X \in \mathbb{R}^+$ , $P_K=\mathbb{P}(X>K) \in [0,1]$ the probability of exceeding it, and $g(x)$: $\mathbb{R}^+ \to \mathbb{R}$, an impact function. Let $G_K$ be a partial expectation of $g(.)$ for the function of realizations of $X$ above $K$:
$$G_K=\int_K^{\infty } g(x) f(x) \, \mathrm{d}x,$$ 
and for clarity lets write the survival function (that is, the complementary cumulative distribution function, CDF) at $K$:
 $$ P_K=\int_K^{\infty } f (x) \, dx$$ 

\begin{figure}[h]
\includegraphics[width=\columnwidth]{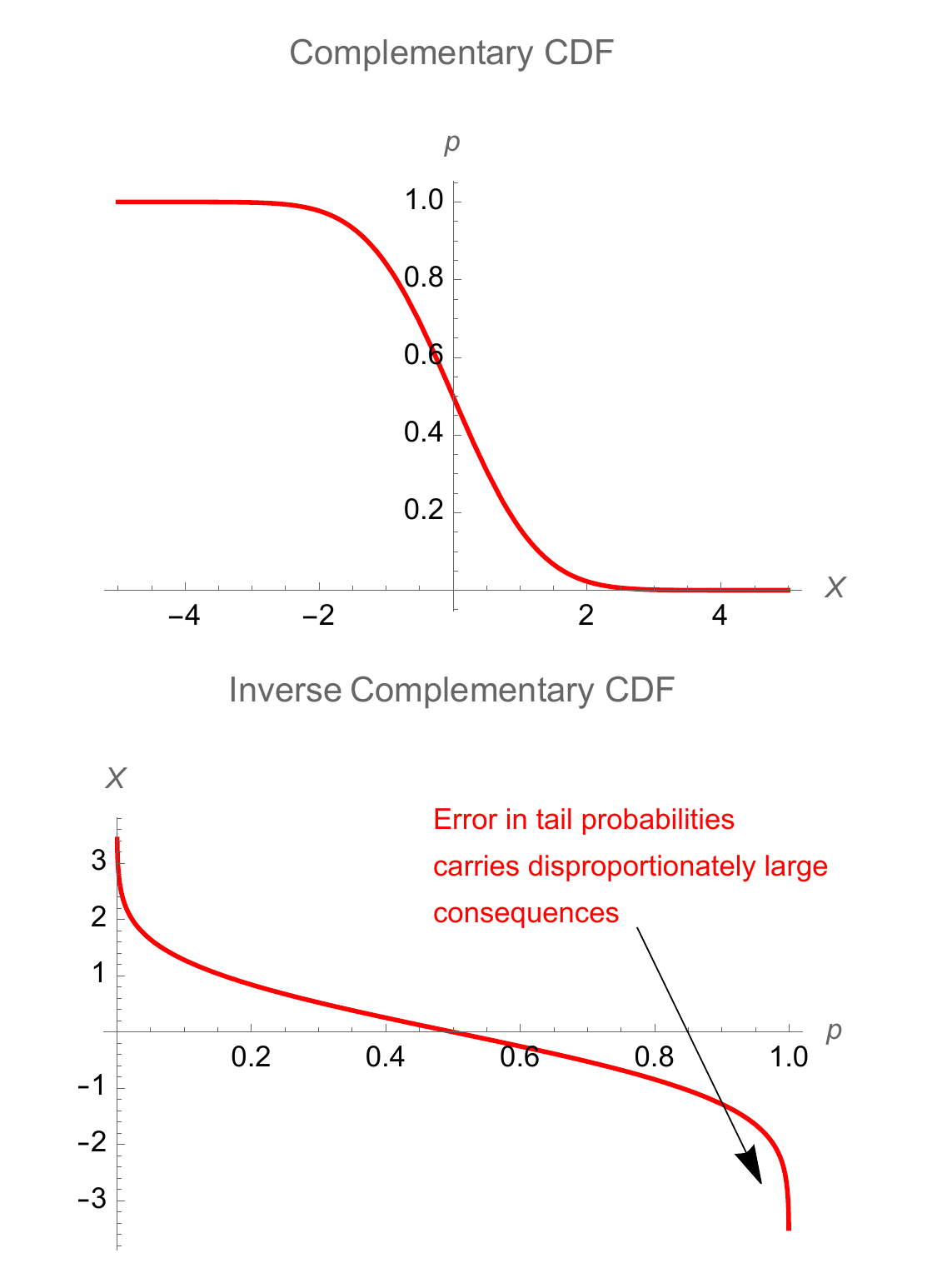}
 \caption{The inverse survival function (complementary quantile function) corresponding to a complementary CDF ($P_K$) is unbounded while the complementary CDF is bounded; it is extremely concave for tail probabilities, and compounds the estimation errors on $p$. Simply the transformation reverses the signs of the second derivative and compounds it. We show how errors on $P_K$ translate in larger and larger values for $K$, possibly infinite.}\label{g1}
\end{figure}

The error comes from conflating the properties of $G_K$ which those of $P_K$, often associating $P_K$ with some constant representing the presumed impact associated with the threshold $K$.

The intuition of the difference can be shown as follows: assuming $g(x)=x$, for $X$ a 
random variable with finite first moment, we have, focusing on the positive domain, generalizing  the Tail Probability Expectation Formula,
\begin{equation}
G_K= \underbrace{K \;P_K}_{\text{Prob times impact at threshold}}+ \underbrace{\int_K^\infty P_x\; dx}_{\text{additional term}},	\label{identity}
\end{equation}
with a second term that can dominate the first, particularly under heavy tailed distributions.

 As explained in the two referenced papers, $P_K$ \textit{as a random variable}, being bounded between $0$ and $1$, necessarily has thin tails, with finite mean, variance, and all moments. By the probability integral transform, its unconditional probability distribution is the Uniform  $\mathcal{U}(0,1)$ --and its conditional is usually treated, particularly in the Bayesian literature, as a Beta (a special case of which is the Uniform) --the Beta distribution can accommodate practically all shapes of double bounded unimodal distributions. A sum of such bets becomes rapidly Gaussian.
\begin{remark}[Moments]
$P_K$ corresponds to the zeroeth moment and is always thin-tailed.	$G_K$ maps to higher moments, and up to the infinite moment (i.e. the extremum) dealt with in extreme value theory.\footnote{Consider the expectation of the $p^{th}$ moment $\mathbbm{E}(x^p)$. $P_K$ corresponds to $p=0$ and the expected maximum to $\lim_{p \to \infty}\mathbbm{E}(x^p)$. In general, risk management concerns extrema, another point of divergence with Tetlock et al.}
\end{remark}

As discussed extensively in Taleb (1997) reflecting the author's experiences as a derivatives market-maker \cite{taleb1997dynamic},  one fails to "hedge" the other in practice (and, of course, in theory). For instance, a rise in skewness of the distribution will tend to increase one side ($G_K$) while decreasing the other ($P_K$): simply, the number of realizations above $K$ drops, but their impact becomes larger.\footnote{Tetlock et al states that Taleb and Tetlock (2013)  claims that the two methods are "complementary". Our understanding of the latter paper is that says the exact opposite: there is no such complementarity fat tailed variables, which is what this discussion about "masquerade" is about. This is the reason why the first author (Taleb) refused to be involved in the IARPA forecasting exercise.} 

And if one does not hedge the other, then being "good at predicting $P_K$" provides \textit{no} information on $G_K$. 

\begin{remark}[Probability Classes]
If $P_K$ and $G_K$ are not in the same probability class, that is, while $P_K$ is always thin tailed, if $G_K$ is not thin-tailed, then one cannot be a \textit{practical} proxy for the other.	
\end{remark}

The field of extreme value theory was designed to deal with the issue. Basically, "probability" is not a tangible object like a tomato; it is, mathematically, a kernel inside an integral (or a summation), inseparable from other integrands, and one should avoid drowning it with verbalism. This point applies no matter the probability interpretation (Bayesian, frequentist, propensional, or subjectivist). Furthermore, science is not about precise measurements of exceedance probability, but understanding properties in a comprehensive and useful ways. As explained in the paper criticized by Tetlock et al (2022), one does not handle pandemics via forecasting tournaments and reliance on champions for single point forecasts, but by getting full distributional properties, particularly the shape in the tails. Decisions must be informed by the shape of the total distribution, and some understanding of the dynamics involved in generating such a distribution (multiplicative effects cause thick tails while additive ones tend to be benign stochastic outcomes).

Just as warning is not predicting, understanding distribution classes and tail properties is not forecasting. Furthermore, the language of "false positive", while useful in medicine and similar applications based on signal, in not useful in risk and insurance based on distributional considerations.

\begin{remark} [Gambling]
It is worth noting that binary options on financial instruments (that is, the trading of $P_K$ or $1-P_K$) proved of little economic value and are not considered an investment in the U.S. and the European Union; they are banned by most corresponding regulators, as they are considered gambling devices. The European Securities and Markets Authority (ESMA) have disallowed retail dealing with binary options.
These binaries were also traded at Lloyds, until banned by U.K. legislation with the 1909 Marine Insurance Act.

\end{remark}

\textbf{Note on "dichotomies":}  Tetlock et al (2022) seem to mix the "dichotomy" between binary and full payoffs with another distinction, that probability estimates help to flag tail risk. Our representation can accommodate both with the function $g(.)$ which as mentioned earlier can reflect the infinite moment, i.e. the extremum.

 \section{A New Result}
 At the Global Uncertainty Reading Group discussion around Tetlock et al (2022), on Dec 1, 2022, a new useful result emerged, which we find worth communicating\footnote{This result can be useful in financial risk management, particulaly the mapping between "VaR" (Value at Risk, maps to $K$ for a set probability $P_K$ of losses above that threshold) and expected shortfall, "CVaR" (maps to $G_K$, that is, includes the impact of losses).}.
 
 \begin{remark}[Events are not defined]
 	A well known problem with heavy tails is that, at the core, in that class of distributions, "events" are not defined verbally: a "war" can have 200 or a million casualties, so it does not have a quantitative meaning. But setting a precise threshold no longer maps to a precise probability under an error rate, and, vice versa. 
 \end{remark}

As illustrated in Fig. \ref{g1}, The error in the evaluation of the probability $P_K$ can translate into an explosive value for the corresponding $K$, and the more fat tailed the distribution, the more explosive such corresponding value.

There is no space for a general proof, so we shall provide one for any distribution that ends (for large values) with Paretan tails, which is the standard case. Let us assume the probability $P_K$ follows a Beta Distribution $\beta(a,b)$ (both parameters $>0$ and as we mentioned this fits the unconditional uniform with $a=b=1$). 
\begin{equation*}
f_{P_K}(p)= \frac{p^{a-1} (1-p)^{b-1}}{B(a,b)}, 
\end{equation*}
$0<p<1$, where $(B.,.)$ is the standard Beta function.

The mean and variance will be $M_{P_K}= \frac{a}{a+b}$, $V_{P_K}=\frac{a b}{(a+b)^2 (a+b+1)}$.

Now assume the underlying distribution for $X$ where $X>K$, lies in the strong Pareto basin, meaning $P(X>K) = L^{\alpha} K^{-\alpha}$, where $\alpha$ is the tail index an $L$ a scaling constant --this is general for large values of $X$ under all fat-tailed distributions.

The inverse complementary CDF (quantile function) can be expressed as:
$ K=L \left(P_K\right)^{-1/\alpha }\text{ if }0\leq P_K\leq 1$.

 If probability taken as a random variable $P_K$, that is, $1-CDF$ (the cumulative distribution function), follows a Beta distribution $\beta(a,b)$, then $K$, the inverse complementary CDF has for density:
 
 \begin{equation}
f_K(k)= \frac{\alpha \,  K^{-a \alpha -1} L^{a \alpha } \left(1-K^{-\alpha } L^{\alpha }\right)^{b-1}}{B(a,b)}
,\end{equation}
with mean 
$$M_K=\frac{L \, \Gamma \left(a-\frac{1}{\alpha }\right) \Gamma (a+b)}{\Gamma (a) \Gamma
   \left(a+b-\frac{1}{\alpha }\right)},$$
   and variance
   $$V_K=\frac{L^2 \,\Gamma (a+b) \left(\frac{\Gamma (a)
   \Gamma \left(a-\frac{2}{\alpha }\right)}{\Gamma \left(a+b-\frac{2}{\alpha
   }\right)}-\frac{\Gamma \left(a-\frac{1}{\alpha }\right)^2 \Gamma (a+b)}{\Gamma
   \left(a+b-\frac{1}{\alpha }\right)^2}\right)}{\Gamma (a)^2}.$$

The proof is done via the standard Jacobian Method for the transformation of probability distributions.

As we can see the first moment exists only if $\alpha>\frac{1}{a}$ and $\alpha>\frac{1}{a+b}$; the second moment exists only if $\alpha>\frac{2}{a}$ and $\alpha>\frac{2}{a+b}$; more generally the $n^{th}$ moment exists only if  $\alpha>\frac{n}{a}$ and $\alpha>\frac{n}{a+b}$.

\begin{remark}[Error Propagation]
	While the error on the probability can be small and controlled, the error on the corresponding quantity under consideration can be infinite.
\end{remark}

We note that the tail index $\alpha$ for pandemics (as addressed at length in the paper criticized by Tetlock et al. (2022)) is well below $1$. The same applies to wars, which means that when it comes to conflicts, forecasts for tail events are not compatible with probability theory.

\section{A Rather Unscientific "Challenge"}

\begin{quotation}
	"We challenge Taleb et al. (2022) to be equally transparent about the performance of their tail-risk hedging strategies —-a controversial topic (Brown, 2020)."
\end{quotation}

We are surprised to see such a remark coming from professional evidence-based researchers: requesting the single track record of a tail hedging program as a back-up for a claim about the mathematical inadequacy of using a binary forecast for, say, Covid 19 or similar events under fat tailed distributions. Said tail-hedging strategy consists in capturing the \textit{difference} between idiosyncratically selected option prices in the market and subsequent market jumps. (Incidentally Professor Tetlock has appeared to conflate tail events --which take place in the tails of  any distribution -- and the fat-tailedness attributes of statistical distributions). And, on top, such request is made to the author of an entire book, \textit{Fooled by Randomness} about the futility of such claims. To repeat the famous disparagement by the economist Jagdish Bhagwati of the claim by the speculator George Soros that he "falsified the random walk" \cite{bhagwati87}, we find it highly unscientific to use a single track record to make any general claim -- this bizarre demand on the part of professional researchers is no different from anecdotal claims ($n=1$) used by medical charlatans. In addition, trading records are not like points in soccer games, particularly when they can hide tail risks.\footnote{Since Tetlock et al (2022) is uncritically citing a web \textit{opinion} article by Aaron Brown, we would like to debunk the claim in it that the performance record is \textit{not} available: not being a retail product, it has been continuously available to what the Security and Exchange Commission (S.E.C.) defines as \textit{qualified} investors, not Twitter activists, and Mr. Brown (who by his own disclosure had a severe conflict of interest) violated, willingly or unwillingly journalistic standards. For it turned out that Brown never did the fact checking, and never asked to see the \textit{audited} returns. We also note that only dimensionless returns are to be compared.}

So we prefer the more robust challenge which, under these circumstances, becomes fair. We believe that academia is about search for knowledge and understanding the world, not a commercial enterprise. The same with societal risk management, which is about the public good and does not issue precise point forecasts (recall that Taleb et al.(2022), as its title indicates, is against single point prediction in some domains). So the burden is on forecasters to forecast.  And we fail to get how a practical project with remarkable forecasting skills could work for government and not the private sector. The superforecasting project is just about such betting. So, as much as we would have preferred to voice the unspoken question, here we are obligated to put the old adage as "if they claim to be so good at forecasting, and their forecasts are actionable and related to reality, why aren't they so rich?" --in other words, why do they depend on taxpayer funds and, possibly, tax deductible (that is, charitable) contributions to finance such forecasts?

\section{Conclusion: Some More Evidence Required}

Finally, can Tetlock et al. prove that better estimates of $P_K$ can provide \textit{real} benefits to decision makers? 

In addition to the problems in the financial domain mentioned above, we completely fail to see the link in insurance --and in event risk in general. In our experience as risk and insurance practitioners, decision makers are usually not well equipped to deal with probabilistic information (compounding the difficulty in translating probabilistic information into practical effects)\footnote{There is the other problem that payoffs are in in expectation space not in frequency space. For instance hedge funds with the best track records turned out to be the most vulnerable to tail events in 2008, see \cite{taleb2020statisticalbook}.}. For instance, in a military context, if we refine the estimate of a South China sea conflict from e.g. 15\% to 17\%, can Tetlock et al prove that this makes a difference in any practical situation?

Also, one is allowed to wonder why the superforecasting project is not applied to sports and election forecasts where 1) $P_K$ applies, 2) compatible with probability theory, 3) provides repeatable tests with overly abundant data and, centrally, 4) is "bankable" (that is, translatable into dollars and cents)?

We conclude with the following recommendation: in future work, it would be helpful if Tetlock et al. provided more rigorous backing of their claims about the link between $P_K$ and $G_K$. For, as it stands, we see neither theoretical nor practical benefits to that "superforecasting" enterprise.

\bibliographystyle{IEEEtran}
\bibliography{/Users/nntaleb/Dropbox/Central-bibliography}

\section*{Supplementary Material}
 The recording of the discussion session (webinar) is available at 
\url{https://www.techincertoreadingclub.com}.

%
%
%
%
%
%
%

 \end{document}